\documentclass[runningheads, citeauthoryear]{llncs}
\usepackage{amsmath}
\usepackage{graphicx}

\usepackage[colorinlistoftodos]{todonotes}
\usepackage[colorlinks=true, allcolors=blue]{hyperref}
\usepackage{pbox}

\begin{document}
\title{Optimizing the Access to Healthcare Services in Dense Refugee
Hosting Urban Areas:\\ A Case for Istanbul}
\titlerunning{Optimizing the Access to Healthcare Services}
\author{M. Tarik Altuncu\inst{1,2} \email{m.altuncu@imperial.ac.uk} \and
Ayse Seyyide Kaptaner\inst{3} \email{ayseyyide@gmail.com}\and
Nur Sevencan\inst{1} \email{nur.sevencan@trtworld.com}}
\authorrunning{Altuncu et al.}
%
\institute{TRT World, Ahmet Adnan Saygun Cad. No:83, Istanbul, Turkey \and
Imperial College London, SW7 2AZ, London, United Kingdom \and
Birkbeck, University of London, WC1E 7HX, London, United Kingdom\\
}
\maketitle              
\begin{abstract}
With over 3.5 million refugees, Turkey continues to host the world's largest refugee population. 
This introduced several challenges in many areas including access to healthcare system.
Refugees have legal rights to free healthcare services in Turkey's public hospitals.
With the aim of increasing healthcare access for refugees, we looked at where the lack of infrastructure is felt the most. 
Our study attempts to address these problems by assessing whether Migrant Health Centers' locations are optimal. 
The aim of this study is to improve refugees' access to healthcare services in Istanbul by improving the locations of health facilities available to them. 
We used call data records provided by Turk~Telekom.

\keywords{refugees\and public health \and access \and migration \and healthcare \and D4R \and CDR}
\end{abstract}
\section{Introduction}

With over 3.5 million refugees, Turkey continues to host the world's largest refugee population~\cite{UNHCR2018}. 
This introduced several challenges in many areas including health sector. 
Refugees have legal rights to free health service in Turkey's public hospitals~\cite{Diker2018}. 
In addition to free health services at public hospitals, 178 Migrant Health centers (MHCs) have been opened in Turkey to fulfill the healthcare needs of refugees and solve overcrowding at public hospitals~\cite{HalkMudurlugu}.
With the aim of increasing healthcare access of refugees, we looked at where the lack of infrastructure is felt the most. 
Therefore, we focus on Istanbul, Turkey's biggest city in terms of population and urbanization, which also happens to be the city that hosts the largest refugee population in Turkey. 
So far, 20 MHCs have been opened in Istanbul. 

Despite the immense efforts for the integration of refugees into Turkey's healthcare system, language barriers~\cite{Bas2015, RePEc:mig:journl:v:15:y:2018:i:1:p:113-124}, registration problems~\cite{RePEc:mig:journl:v:15:y:2018:i:1:p:113-124, Bahadir2015}, navigation of the system~\cite{RePEc:mig:journl:v:15:y:2018:i:1:p:113-124}, overcrowding~\cite{Ozdogan2014, Savas2016, doi:10.1080/20477724.2017.1349061}, refugees' lack of knowledge about the services available to them~\cite{Diker2018}, and lack of translators ~\cite{Diker2018} remain as the main challenges. 
The studies cited above point to challenges refugees face in Turkey . 
There are not extensive studies focusing on refugees' health access in Istanbul. 

In order to check whether Turkey-wide problems facing refugees also apply to refugees in Istanbul, we obtained data from Sultanbeyli Municipality's Refugees Association through Syrian Coordination Center Software (SUKOM). 
According to the data SUKOM provided, only 670 (50\%) of the refugees who reached out to the Association made appointments for the local public hospital in Sultanbeyli. For
instance, out of the 1332 refugees who contacted the association to request language assistance for making a hospital appointment, 360 refugees requested an appointment at a public hospital in Pendik, a further away district in addition to other appointment requests made for hospitals in other districts such as Uskudar, Umraniye, Atasehir, Levent, etc.
Making appointments in districts other than Sultanbeyli can be explained by these two scenarios  which point to either to overcrowding or lack of language support:
\begin{enumerate}
    \item  Not all the refugees who reach out to SUKOM for language assistance reside in Sultanbeyli, thus what we describe as further away hospitals could be their local hospitals or hospitals closer to them 
    \item Overcrowding at the Sultanbeyli public hospital can be expected since it is the district with one of the highest refugee populations in Istanbul. 
\end{enumerate}

501 out of 639 refugees who made their appointment between February 2018 and September 2018 through the association were also accompanied by a translator from the association during their hospital visit. 
This illustrates a lack of language support at hospitals.

Our study attempts to address the problems of lack of language support and overcrowding by assessing whether the Migrant Health Centers' locations are optimal. 
The aim of this study is to increase the access of refugees to healthcare services in Istanbul by improving the locations of health facilities available to them by using the cellular network usage data provided by Turk~Telekom.

We tested if Migrant Health Center (MHC) locations are optimal. 
We suggested optimal locations for Migrant Health Centers.

We considered two different methods of optimization for MHC locations: distance minimization and
duration minimization. According to our results, the locations  we suggest
are more cost effective than their current locations.

\section{Methodology}

Call Detail Records(CDR) are being used by researchers for different purposes such as urban planning, anomaly detection, and understanding people's behaviour under certain circumstances.
A small portion of the literature also consists of network science approach trying to model complex social systems in~\cite{Zhi-Dan2008a}, and understanding people's movement patterns for transportation planning ~\cite{Demissie2016a}. 
Here we use a similar data source to optimise access to healthcare services of refugees living in Istanbul, Turkey.

\subsection{Datasets}

We use the mobile telecommunication datasets provided by Turk~Telekom's Data for Refugees competition~\cite{Salah2018a}. 
The datasets consists of geographical lookup tables for each cell tower (BTS) and other geographical identifiers, call and text communications among each BTSs, call and text activity details of a sampled set of subscribers with their connected cell tower, refugee statuses of callee and callers, and another dataset of a larger and more comprehensive sample of users but with reduced resolution of details.
In this study, we only use call activity details of refugees (we will refer this data as \textit{CDR} or \textit{CDR data}) together with their corresponding geographical locations of cell towers (we will refer this data as \textit{BTS records}).

Although a more comprehensive description has been published by the organizers of the competition~\cite{Salah2018a}, we provide a short description of the datasets below.
The BTS records contains each Turk~Telekom cell towers' geographical coordinates, city and borough details along with unique IDs which we use to join with the CDR data in order to detect geographical coordinates of calling activities.

The CDR data also contains the following information;
\begin{itemize}
    \item caller ID with two categories (refugee or non-refugee),
    \item whether the receiving-party of the call is a refugee or non-refugee,
    \item whether the call is an inbound or an outbound call with reference to the caller ID,
    \item the timestamp for the call made,
    \item the ID of base stations to which the caller is connected at the beginning of call.
\end{itemize}

\subsection{Computational Approach and Data Pipeline}
\label{subsection:pipeline}
Our methods create a data pipeline starting with the detection of the residential location for each refugee subscriber provided in the CDR. 
We derive the residential location for a subscriber based on their night time calling activities registered in the CDR.
Aggregating the residential locations of these refugees in the form of total number of refugee residents per cell tower enables us to determine refugee densities at each site, and to illustrate this on a Voronoi region\footnote{Voronoi regions are polygonal regions constructed by unit areas that
have the same number of base stations as the nearest one~\cite{Zhi-Dan2008a}.} map~\cite{Voronoi} of Istanbul. 
As there exists an excessive number of cell towers in the city, we cluster them on the basis of their proximity to each other and the number of refugee residents in their respective Voronoi regions.
The centers of these clusters constitute the centers of residential regions for refugees.
We obtained the distance and the duration of using public transit among residential region centers from a commercial API. 
Then, we computed the optimum locations for the migrant health centers.
The rest of this section provides step by step detailed information about the pipeline.

\subsubsection{Data Adaptation}
We made some changes in the datasets before starting computations.
Along with some warnings mentioned in \cite{Salah2018a}, we experienced some other problems and inconsistencies within the datasets. 
For the purpose of reproducibility of the results, we describe the changes we made on the data.
\paragraph{BTS Records:}
At first we dropped the BTS records with no geographical coordinates provided due to our need for higher geographical precision than city level resolution. 
Then we converted degree, minutes, seconds (DMS) syntax to latitude and longitude based coordinates.
Further, we merged some BTS records because they were either too close to each other to distinguish in terms of the region they cover or on exactly the same coordinates. 
We used DBSCAN \cite{Ester1996} algorithm with epsilon value of 0.0005, and using the euclidean distance metric to measure the distance between geographical coordinates. 
While joining the BTS with CDR, we used inner join method in order to avoid records with lack of either call details or geographical details. 
Finally, we discovered some BTSs have wrong city records. 
For instance, some BTSs that are supposed to be in the city of Bolu had geographical coordinates which are actually in Istanbul. 
Hence, we did not use the provided city information of BTS records. 
Instead, we mapped all BTSs and filtered out the ones that fall outside of Istanbul's official district boundaries. 
We also use the same method to determine whether the cell tower is located in European or Asian side of Istanbul. 
This feature is further explained in \nameref{section:bts_clustering} section.
We mirrored all changes and filters in BTS records to the CDR data using the \textit{'SITE\_ID'} field.

\paragraph{CDR:} 
We filtered CDR data to cover only the voice calls made by the refugees and made in Istanbul using the merged BTS coordinates. 
To note, we did not rely on the \textit{'CALL\_TYPE'} flag on any part of our analyses because we saw that many users are making either only inbound or only outbound calls according to this feature. We believe that this phenomena is probably a result of sampling made by the data provider, and does not reflect the subscribers' behaviour.

\subsubsection{Computing the Number of Refugee Residents}

Detection of a user's residential region via cellular network usage is an ongoing attempt as various form of approaches can be found in the literature. In \cite{cdr_gprs_home_work}, authors attempt to detect home location based on activities made nearest to sleeping period which is approximated using the inactive time periods per user, whereas in ~\cite{10.1007/978-3-642-21726-5_9} a small subset of labelled data was used to train a logistic regression model to detect important sites based on the activities made during \textit{'home hours'}, which they define as being between 7 PM and 7 AM.
Although the limitations of such methods apply to many cell-phone users having unusual call and text patterns or low usage at individual level, it still provides a quantitatively reliable measure at aggregate level when there is sufficient data.

Similar to~\cite{10.1007/978-3-642-21726-5_9}, we make an assumption that people are active at their homes during the nighttime, but we selected a narrower time interval which is between 11 PM and 8 AM. 
Among all users who have ever registered in Istanbul, 62\% of all subscribers and 54\% of refugees have at least one call records in this time interval.
Using the number of calls which the refugees make during the nighttime, we extract the frequencies of cell towers being used per refugee subscriber. These frequencies can be regarded as the probabilities of corresponding user's residential address in the area covered by the cell towers. To detect the boundaries of regions served by cell towers, we calculate Voronoi regions for cell towers and refer to these regions as \textit{cell tower areas}.

Since we are interested in finding the refugee distribution in Istanbul, for each cell tower area we sum up the probabilities of refugees living there. 
This new way of representing the same computation gives us the advantage of mapping refugee density using choropleth maps. 
However, this map creates a very fine resolution of Istanbul since we have more than 4 thousand cell tower locations based on our BTS dataset. 
As the scope of our objective is finding the optimal placement for MHCs, we will not use this resolution in our analyses.

\subsubsection{Observation Weighted Clustering of Cell Towers}
\label{section:bts_clustering}

To obtain coarser grain representation of the Voronoi cell regions, we apply k-means clustering~\cite{Arthur} to each cell tower. 
Although this algorithm is directly applicable to spatial positions in two dimensions, our cell towers are not identical in terms of their inertia because they have varying number of refugees. 
In line with our objective of providing higher resolution for refugee dense regions but coarser resolution for the rest, we weighed each cell tower's location with its corresponding number of refugee residents. 
To compute this, we modified the input to k-means algorithm in a similar fashion described in the referenced blog post~\cite{Anderson2016}.

For k-means clustering algorithm one has to provide the number of clusters as \textit{a priori}. For our case, we set total number of clusters to 200. This is an intuitively selected quantity which is intended to be small enough to be comparable to the number of MHCs while being large enough to maintain enough resolution for refuge-dense regions 

We have an additional concern about the geographical boundary of the Bosporus which divides Istanbul into two continents. 
This natural boundary creates bottle-necks in the transportation networks which local people mostly avoid although there are multiple transportation options that could potentially ease the commute only from specific locations.
Therefore, we know that although the two coasts of the Bosporus are very close, we have to avoid grouping
regions from different continents together as the transportation between the two is generally
more time-consuming. 
Therefore, we divide the cell towers into two sets for Europe and Asia. 
Then we separate the parameter k into two parts based on the proportions of refugee call activities from each side of the Bosporus using the CDR data. 
Since the Asian side consists of only the 35\% of all calling activities, we assign 70 clusters to Asian side. 
We assign the rest 130 clusters to European side.

\begin{figure}[h]
\includegraphics[width=\textwidth]{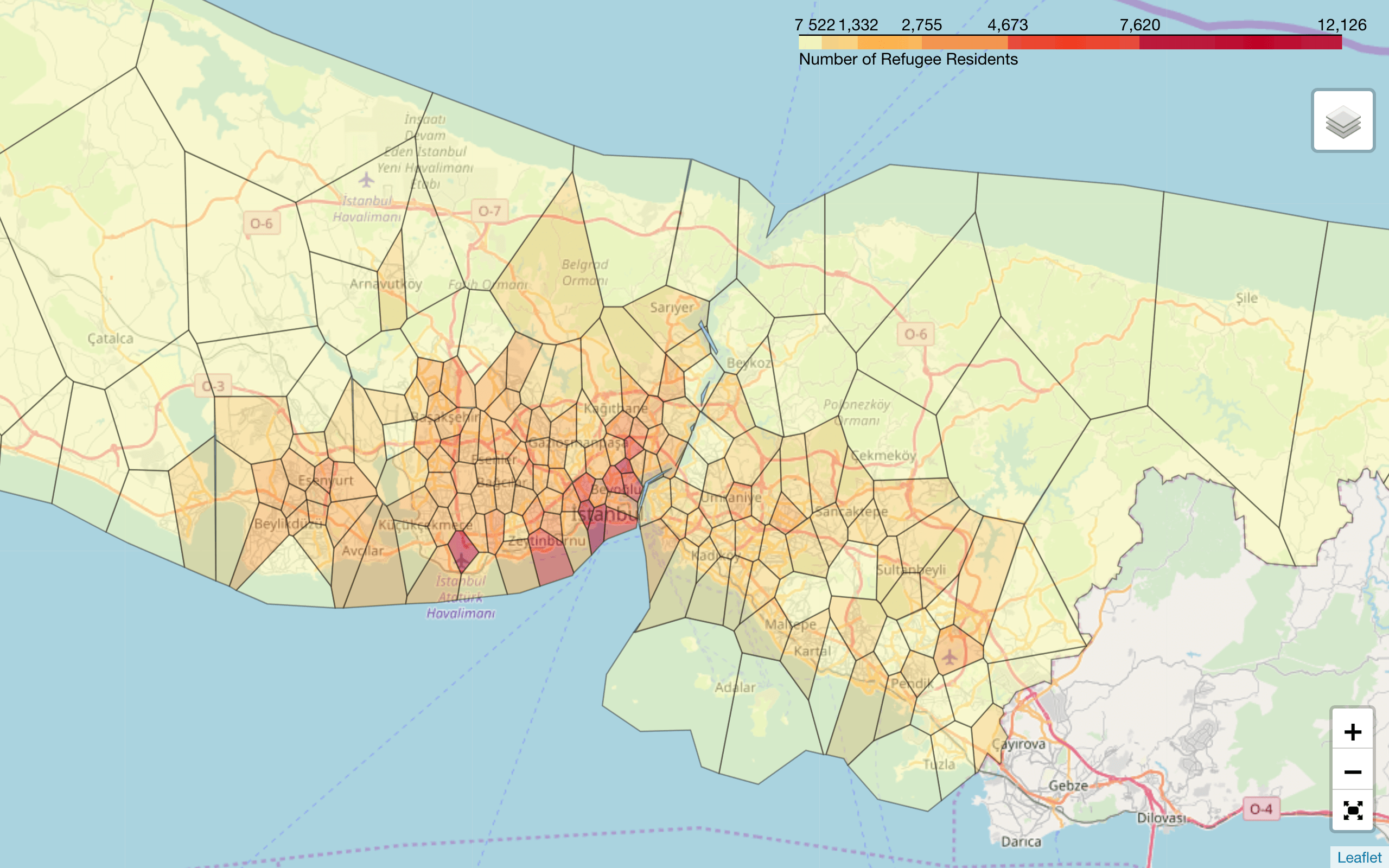}
\caption{Map shows the number of refugee residents per residential region in Istanbul based on the number of refugee residents metric which we computed using the night time calling activity of refugees in CDR data.} \label{fig:regions}
\end{figure}

After running two k-means clustering algorithm for two sides, we assign our residential region centers to the central points of k-means clusters. 
We create another set of Voronoi cell regions using residential region centers and refer them as \textit{residential regions}.
Lastly we add up the refugee residents in all the cell towers in each cluster and plot them in Figure~\ref{fig:regions} as choropleth map.

\subsubsection{Transit Costs Between Clusters}
\label{section:googleapi}

Having 200 clusters as residential regions and their centers, we must obtain the average costs of public transit among these locations before running any optimization. For that, we used Google Cloud Platform's Distance Matrix API\footnote{Details about the Distance Matrix API could be obtained from \url{https://developers.google.com/maps/documentation/distance-matrix/start}}. We requested its estimated time and distance costs between pairs of all residential region centers at 10 AM with local time on December $10^th$, 2018 via public transit. We converted the results into two matrices where $D$ contains distances in meters and $T$ the transit duration in seconds. For some cells that Google failed to provide, we used the mean value of the rest of the matrix.

\subsubsection{Optimizing the Access to MHCs}
\label{section:optimization}

We applied a multi-facility location optimization linear programming model based
on the problem described and solved in~\cite{ReVelleCharlesS1970CFL}. The problem considers selecting m
locations from n candidates, considering either the distance or the duration matrix
between all pairs of n candidate locations with the objective function of minimizing the
total travel costs of various number of people initially located in n locations.

The equation for this problem is given below;

\begin{equation}
\begin{aligned}
    \underset{x}{\text{minimize}}
    & \sum_{i=1}^{n}\sum_{j=1}^{n}a_i \cdot d_{ij} \cdot x_{ij} \\
    \text{subject to}
    & \sum_{j=1}^{n} x_{ij}=1, \; i = 1,2,\ldots,n, \\
    & x_{jj} > x_{ij}, \; i = 1,2,\ldots,n, j = 1,2,\ldots,n, i \neq j, \\
    & \sum_{j=1}^{n} x_{ii} = m \\
    & x_{ij} > 0, \; i=1,2,\ldots,n, j = 1,2,\ldots,n.
\end{aligned}
\end{equation}

\begin{figure}[h]
\includegraphics[width=\textwidth]{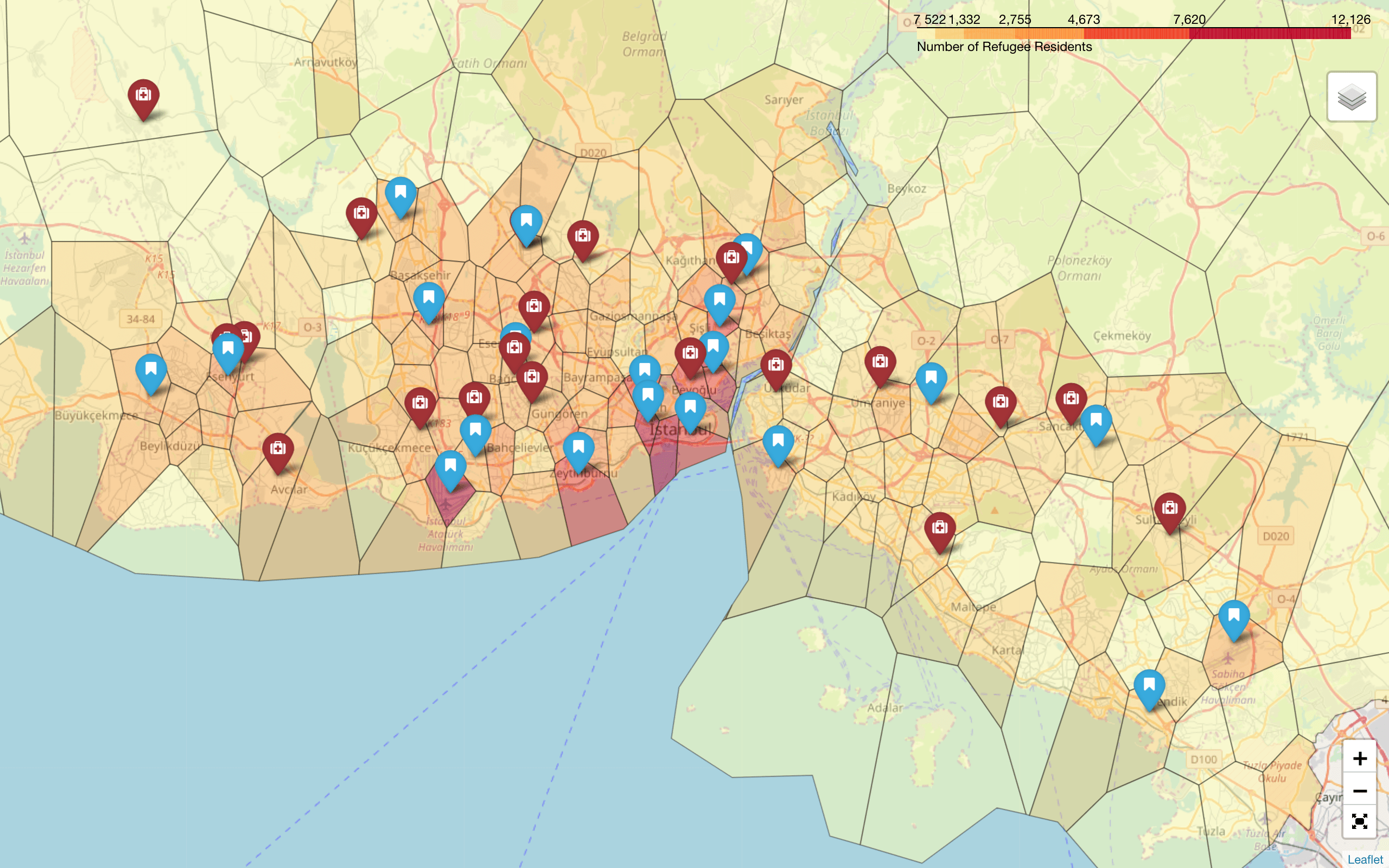}
\caption{Map with all current MHCs (red pins) and optimized MHC locations based on travel durations (blue pins) over choropleth map with the refugee residential densities.} \label{fig:mhcs-duration-1}
\end{figure}

Where $a_i$ is the number of people initially located at point $i$, $d_{ij}$ represents the distance or transit duration matrices for travelling from point $i$ to point $j$, and $x_{ij}$ is the decision variable which we optimize using PuLP optimization package~\cite{Mitchell11pulp:a} in Python. 
The solution matrix $X$ assigns $n \times n$, and assignment is indicated by the entry $1$ for each row on its assigned column and 0 for the rest.
Array input $a$ consists of the number of refugee residents for all residential regions. 
We set $n$ to 200 as we have 200 candidate locations, and $m$ to 20 because Istanbul currently hosts 20 MHCs at the time of our analysis. 
The optimization has been run twice because we found optimum MHC locations with both using the distance $D$ and transit duration $T$ matrices. Figure~\ref{fig:mhcs-distance-1} and Figure~\ref{fig:mhcs-duration-1} shows the produced results respectively.

\section{Results}

Our results are also plotted on an interactive map\footnote{The interactive map is published on \url{http://bit.ly/refugee_map}}.
On the map, the dropdown menu displays the following layers: Cell Tower, Cell Tower Areas, Residential Regions, Residential Region Centers, Current MHC Locations and MHC Locations Optimized by Distance/Duration. \textit{Cell Towers} are where calls and texts are first registered. \textit{Cell Tower Areas} shows the regions the cell towers cover. As explained in \nameref{subsection:pipeline} section in detail, we grouped similar cell towers based on their proximity to each other and where refugees make phone calls during the night time. The centers of these cell tower regions are referred to as \textit{Residential Region Centers} which later constituted the Voronoi cells which are referred to as \textit{Residential Regions} in our study.

\section{Discussion}

The availability of migrant health centers improves refugees' access to healthcare. We focused on
location optimization because healthcare literature finds negative correlation between the distance travelled to hospitals and health outcomes \cite{Kelly2016}.

\subsection{Proposing Optimal Locations for Migrant Health Centers}
CDR data provided by Turk~Telekom enabled us to create an alternative
residential map of refugees. Refugee demographics in Istanbul are not
registered in a census data. Instead, each municipality keeps refugees' information
in their database. Their address changes might not be recorded
especially given that refugees move often as they settle down in the host cities.
By taking this into account, we initially took the locations where refugees spend the most time (their
mode location) as their departure points for hospitals. However, we saw that it does
not really match with the refugee distribution in Istanbul as seen in the study by \cite{ISTalmanak15}

\begin{figure}[h]
\includegraphics[width=\textwidth]{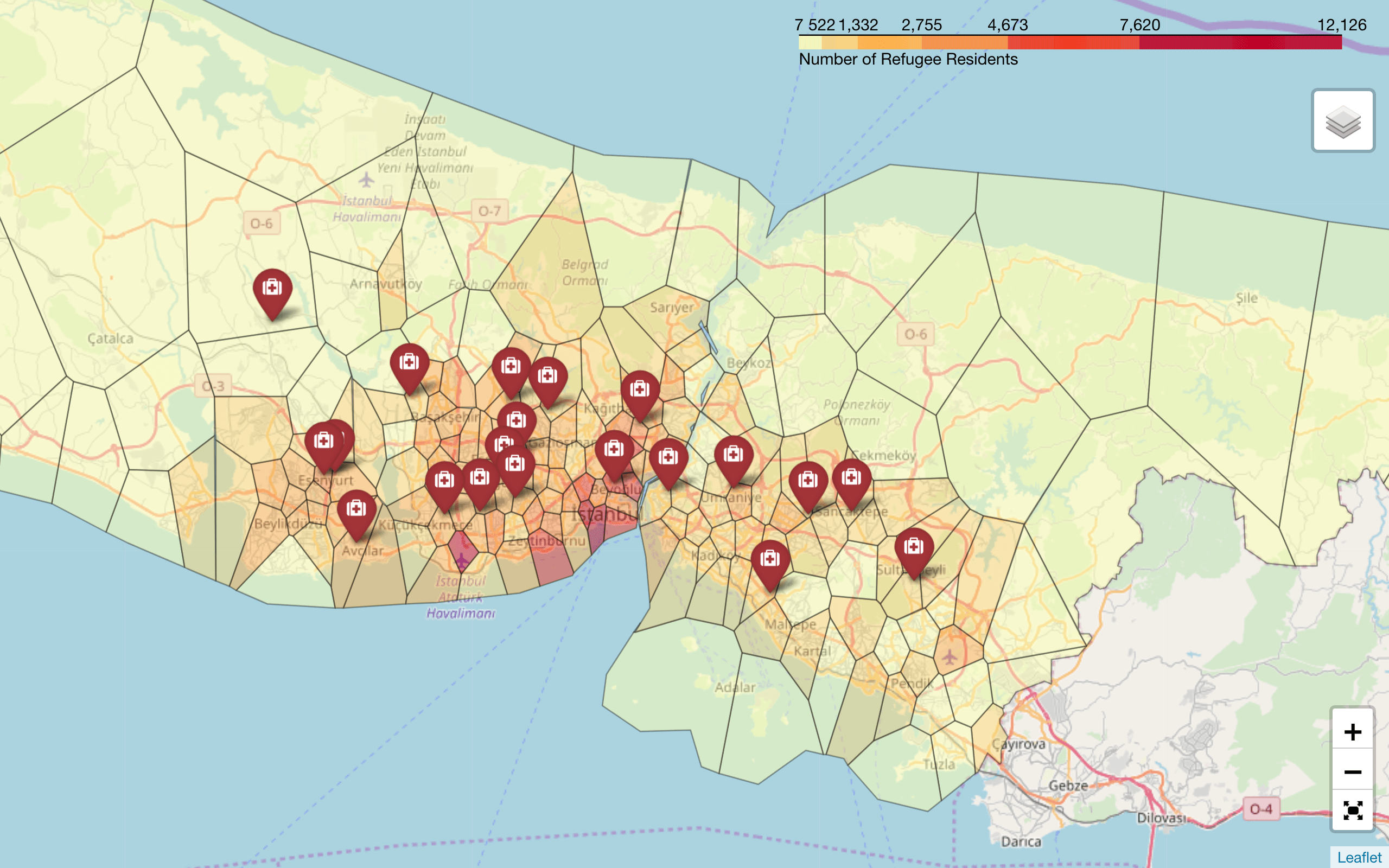}
\caption{Map with red pins showing the locations of current MHCs over the choropleth which illustrates the density of refugee residents at different districts in Istanbul.} \label{fig:current-mhcs}
\end{figure}

\begin{figure}[h]
\includegraphics[width=\textwidth]{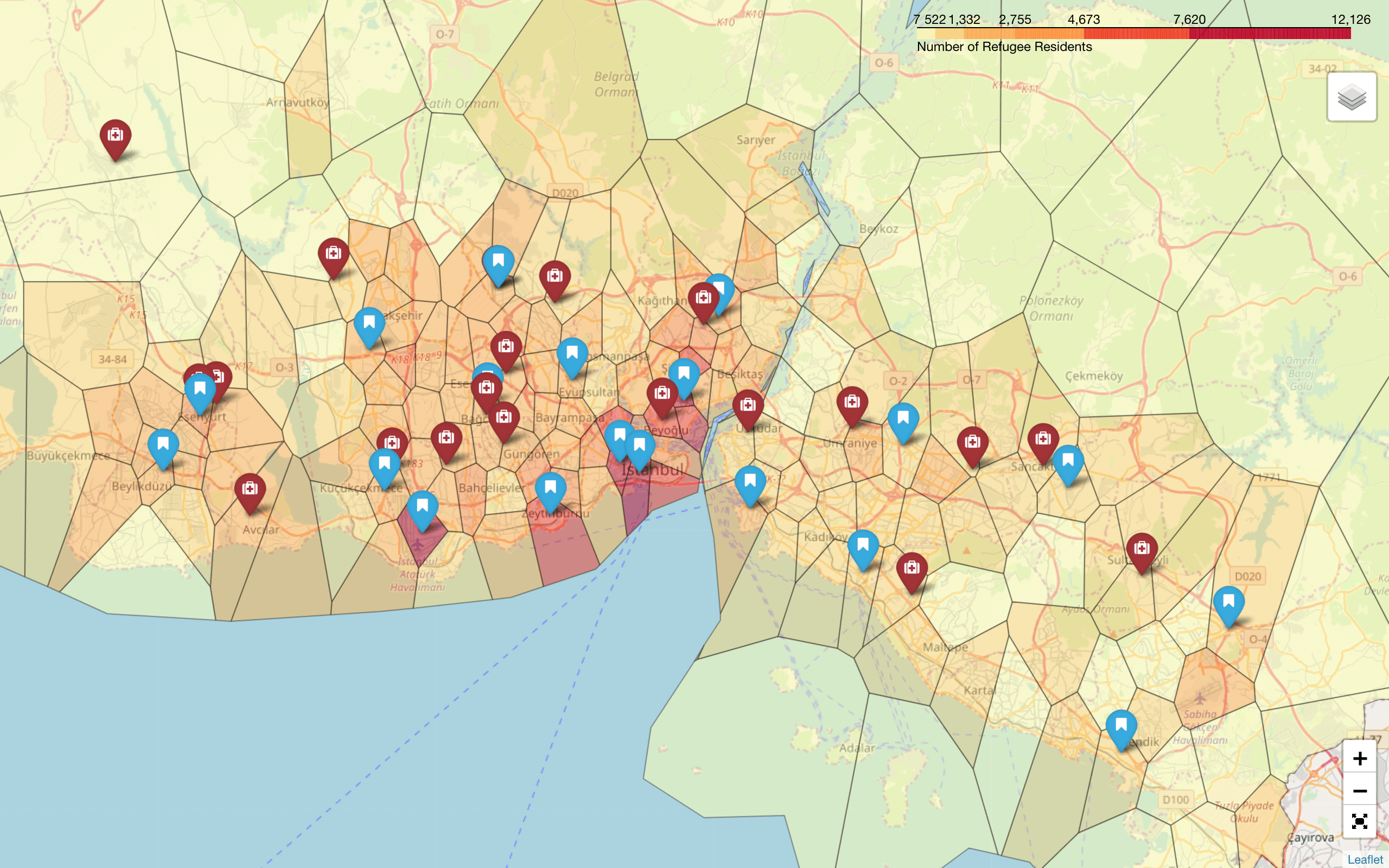}
\caption{Map with all current MHCs (red pins) and optimized MHC locations based on travel distances (blue pins) over choropleth map with the refugee residential densities.} \label{fig:mhcs-distance-1}
\end{figure}

Our results suggest that the mode locations are more likely to be places where refugees
socialize and spend their daytime regularly, such as the Aksaray district. Thus, using their home
address is a better proxy for the access available in each municipal district.
However, modes we identified through our study can be used to find optimum locations for other public services, such as social integration centers as
well as employment and education facilities.
Our method of estimating refugees' home addresses using CDR data is explained in detail
in our methods section. We calculated optimal locations for the Migrant Health
Centers (shown with blue pins on Figure~\ref{fig:mhcs-distance-1}) and compared them with the existing
MHC locations (indicated by red pins on Figure~\ref{fig:current-mhcs}). The MHC locations our optimization suggest are not in exact coordinates, they can be positioned at any central place around the pinned region.

\begin{figure}[htb]
\includegraphics[width=\textwidth]{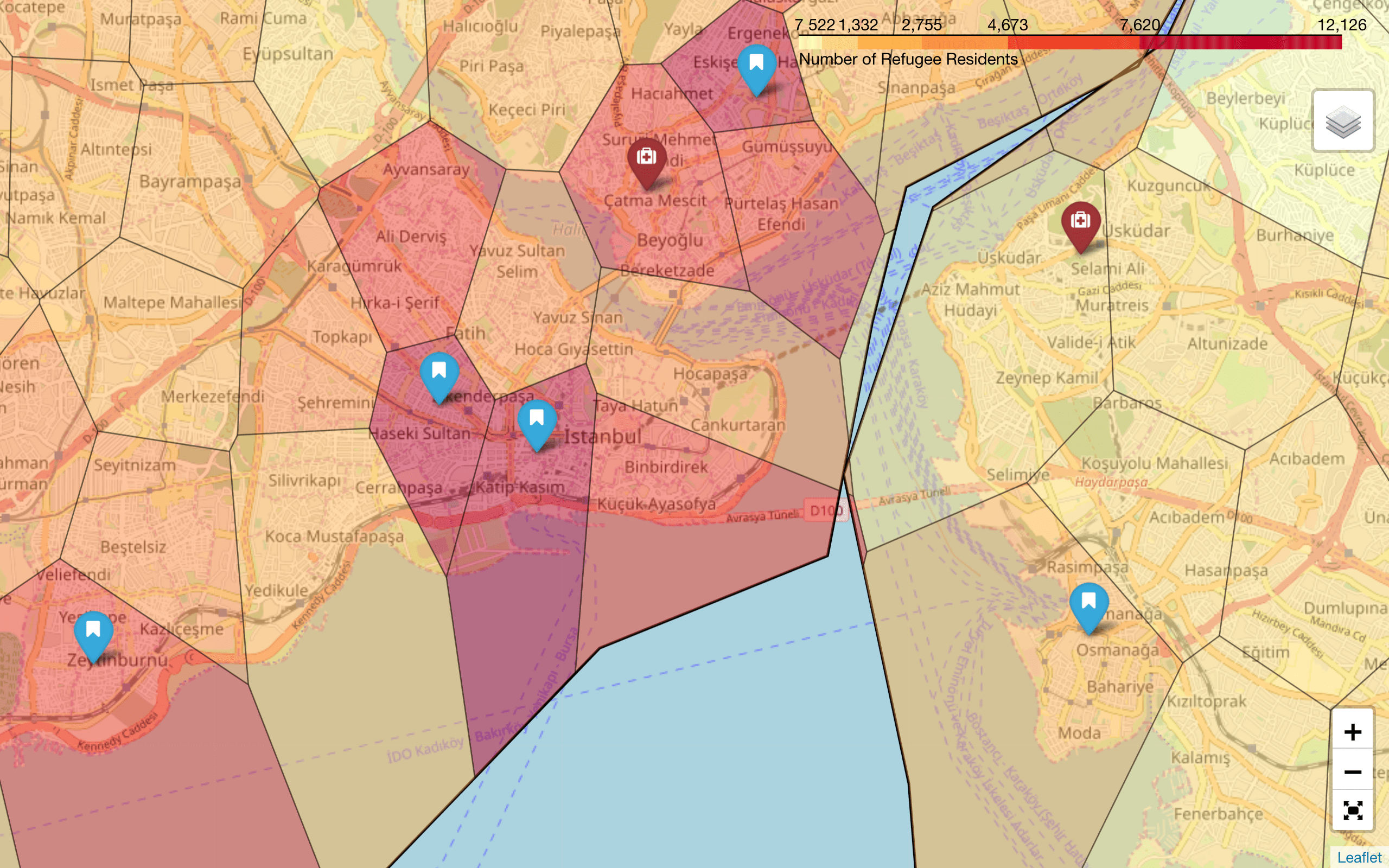}
\caption{A close-up view of central Istanbul where optimized locations based on travel distances (blue pins) show particular districts where new MHCs can be opened} \label{fig:mhcs-distance-2}
\end{figure}

As can be seen in Figure~\ref{fig:mhcs-distance-1}, the locations we suggested are sometimes very close to the already existing MHCs. Notable differences can be better observed in Figure~\ref{fig:mhcs-distance-2}). For example, our optimization suggests MHCs in Yeşilköy near Ataturk Airport (lower left) and near the Sabiha Gokcen Airport(middle right). We believe this is caused by the continuous call traffic in the airports that are mistakenly identified as refugee residences by our resident detection method. Another noticeable difference between optimal and current locations can be observed around historical peninsula(Fatih district) in Figure~\ref{fig:mhcs-distance-2} . Currently, there is no MHC in the area, whereas our optimization suggests locating three MHCs there. We believe this to be a crucial insight since MHCs can provide refugees with primary and secondary healthcare services without facing language barriers and overcrowding at public hospitals.

Travelling costs in Table~\ref{tab1} are the sum of travelling distances and durations for all refugees in the case that each refugee visits their nearest MHC averaged by the total number of refugees who we can detect their residencies. Cost of travel distances and durations are shown to represent an average cost for one refugee to access their nearest MHC for three cases: Current MHC locations, distance based optimized MHC locations and duration based optimized MHC locations. 
Besides the current MHC locations, we present the overall performance of our suggested MHC locations which have been optimized to minimize the travel cost on the basis of distance and duration.

Overall, we found that the average travelling cost to MHCs is 5.9 km for one person and it takes approximately 26 minutes in the present case. 
The locations we proposed based on refugees' activities from CDR data cuts the travel distance down to as little as 3.6 km in distance based optimization, and travelling time to as little as 18 minutes for duration based optimization.

\begin{table}
\centering
\caption{Cost of travel distances and durations for current MHC locations and optimized MHC locations (on the basis of distance and duration).
All distances are provided in kilometers and durations in minutes.}
\label{tab1}
\begin{tabular}{|l|c|c|}
\hline
MHC Locations & \pbox{20cm}{Travel Distance} & \pbox{20cm}{Travel Duration} \\ \hline
Current &  5.9 & 26 \\ \hline
Optimized (Distance) &  3.6 & 20 \\ \hline
Optimized (Duration) &  4.4 & 18 \\ \hline
\end{tabular}
\end{table}

\section{Conclusion}

One of our biggest challenges during this study was the lack of health care service data
available for public use.
Another major limitation we can define for this study is the representativeness of the
sample. The CDR data shows that 85\% of the total activities of refugees in Istanbul
are in Europe, while 15\% is in Asia. We suppose that this discrepancy could be because of the data's inherent bias which is that the refugee population which uses Turk~Telekom does not represent the
whole refugee population.

Our paper proposes the following:
\begin{itemize}
    \item Opening Migrant Health Centers by optimizing travel time, in Istanbul's
case in districts such as Zeytinburnu, Fatih, Kadikoy, and as shown in Figure~\ref{fig:mhcs-distance-2}
    \item Our study focuses on improving access to healthcare services for refugees
in Istanbul. However, our approach is applicable and scalable to other
cities.
    \item When it is not possible to open a new MHC or move existing MHCs to more
optimal locations, translation services that will help refugees with making
appointments and communicating in hospitals can be offered near the
optimal locations we proposed.
    \item We strongly encourage that refugees are directed to public hospitals instead
of research or university hospitals that are already overcrowded.
\end{itemize}

Despite the limitations of the CDR Dataset and the lack of supporting datasets, our estimation of refugee residence locations is possibly more robust than official records given that refugees can be very mobile when they first come to a new country and they can neglect to update authorities with their change of address, making it difficult for Turkish authorities to track their up-to-date address. If other cellular network providers in Turkey share their CDR data, the accuracy of location optimization would be significantly higher.

This study hopes to bring a new perspective on determining the locations of healthcare facilities. Further research on the availability of necessary healthcare units and healthcare staff in MHCs, language support in public hospitals, and the accessibility of hospitals for special treatment needs of refugees are needed.

%
%
%
\bibliographystyle{splncs04}
\bibliography{mendeley_d4r}

%



\paragraph{Abbreviations}
MHC: Migrant Health Center
BTS: Base Transceiver Station
CDR: Call Detail Record
SUKOM: Syrian Coordination Center Software

\paragraph{Author\'s Contributions:} Muhammed Tarık Altuncu conducted the computational
research. Nur Sevencan and Ayşe Seyyide Kaptaner were responsible for the
social science research and obtaining additional data. All authors analysed the
data and wrote the manuscript.

\paragraph{Acknowledgements:} We thank the rest of the members of TRT World's team for
the Data for Refugees challenge: Basri Ciftci, Berk Baytar, Soud Hyder, and
Yasin Sancaktutan. We also thank Mehmet Efe Akengin for providing intellectual support,
and Hamza Osmanogullari for logistic support.

\end{document}